\pdfoutput=1
\documentclass[10pt,a4paper,notitlepage]{revtex4-1}
\usepackage{graphicx}
\usepackage{amsmath}
\usepackage{subcaption}
\usepackage{color}
\usepackage{fullpage}

\begin{document}
\title{Frequency shifts of radiating particle moving in EIT metamaterial}

\author{S. Zieli\'nska-Raczy\'nska}
\author{D. Ziemkiewicz}
\email{david.ziemkiewicz@utp.edu.pl}
\affiliation{Institute of Mathematics and Physics, University of Science and Technology, Al. Kaliskiego 7, 85-798 Bydgoszcz, Poland.}

\begin{abstract}
 
Nowadays, there is considerable interest in metamaterials which realize the electromagnetically
induced transparency in a classical system. 
We consider the frequency shifts of particle moving in  metamaterials exhibiting electromagnetically induced transparency effect. The dramatic change of the material dispersion due to the EIT  influences the conditions for signal propagation in the medium and has a significant impact on the Doppler effect, possibly leading to the optical control over this phenomenon. The dependence of the Doppler shift to the source frequency and velocity  and radiation spectra  on external parameters is examined. It was found that for source frequencies fitting transparency window for particular range of source velocities cutoffs appear, i.e. the radiation is not emitted. Our theoretical findings are proved analytically and confirmed by numerical simulations based on finite-difference time-domain method.

\end{abstract}
\maketitle 
\section{Introduction}

Electromagnetically induced transparency (EIT) is a well-known phenomenon of quantum interference which bases on extraordinary dispersive properties of an atomic medium with three active states in the $\Lambda$ configuration.
 This phenomenon leads to the reduction of absorption of a resonant probe laser field by irradiating the medium with a strong control field making an otherwise opaque medium transparent \cite{Boller}- \cite{Fleisch}. EIT is primarily concerned with the modification of linear susceptibility, which requires explanation based on quantum interference phenomena; the dark state i.e. the state of zero absorption, is caused by destructive interference.
EIT serves to drive atomic coherences and involves the destructive quantum interference between different resonant excitation pathways of atoms; it leads to dramatically changes of dispersion properties of the system -   absorption  forms a dip, called the transparency window, and approaches zero while dispersion in the vicinity of this region becomes normal with the slope which increases for a decreasing control field - the resonant probe beam is  now transmitted almost without loses. Since at least twenty years there has been a considered level of activity devoted to research of EIT. This has been motivated by a recognition of a number of its applications among which storing the light in a medium is a well-known example \cite{LiuEit},\cite{Fleisch2}.

During the last years a lot of work has been done to study, both theoretically and experimentally more and more sophisticated variation of EIT \cite{Fleisch}. By admitting atoms with more active states, coupled by more control fields in various configurations it has become possible to dynamically change the
 optical properties of more complex media \cite{Racz1}-\cite{Racz4}
and by optical means reveal new aspects of quantum optics which could be applied to constructing efficient tools for photonics, e.g. quantum memories, switches, multiplexer.

Extensive studies of EIT in an atomic media inspired the interest of application of this phenomenon in others media.  It turned out that  plasmonic molecule shows electromagnetic response that closely resembles EIT in atomic media \cite{Zhang}.
Due to the considerable practical usefulness of EIT a lot of efforts has been devoted to develop classical analogies of this phenomenon and related effects. Nowadays, there is a great interest in creating a new class of specifically structured metamaterials in which a classical analogue of electromagnetically induced transparency could be realized \cite{Souza}, \cite{Tassin}. Such systems follow the same dynamics and allow for manipulation of the macroscopic media properties resulting in tunability of EIT. The investigations of classical analog of EIT media have been motivated by recognition of a number of potential new applications in such media: lossless propagation of a signal through initially thick media, wide bandwidth, high-efficiency nonlinear optical processes and in addition operation at room temperature.
EIT, reversing the dispersion properties of the medium, changes drastically the conditions for signal propagation in the medium i. e. initially opaque medium exhibits almost perfect transparency in a vicinity of certain (resonant) frequency, allows for pulse propagation in otherwise optically thick medium  and inside the medium the group velocity of the signal becomes significantly altered.

It has been shown in \cite{Ziemkiewicz} that the material dispersion has a significant impact on the Doppler shift and leads to the complex Doppler effect where monochromatic source generates multiple frequency modes. Moreover, the group velocity of the these modes play a central role in the formation of the radiation pattern around the source \cite{Ziemkiewicz2}. In particular, a phenomenon similar to the Cherenkov radiation occurs for sufficiently small group velocity. Therefore, it might be interesting to investigate the phenomena associated with moving radiation sources in a medium where EIT is realized,
possibly leading to a new class of devices where these effects might be optically controlled and propagation dynamic could be changed on demand.

This paper is devoted to study of the Doppler effect in an arbitrary system in which EIT is achieved. When the source of  radiation enters such medium, due to so-called complex Doppler effect, one can detect multiple modes of shifted frequency   \cite{Ziemkiewicz}. We consider two situations: first, when the frequency source fits transparency window, and second when radiating source enters EIT medium with frequency out of the window. We give relatively simple analytic formulas which allow one to explain the nature of frequency changes. We report the shifts dependence on various system parameters. Recently, Bortolozzo et al \cite{Bortolozzo} have proposed the robust, balanced detection method to ultraprecise and stable Doppler shift measurements which could be performed in various systems (i. e. slow light media or acoustic systems) with steep dispersion. In such systems even small Doppler shift unbalances the two output intensities. Therefore our findings could be helpful in analysis of precise Doppler shift measurements in a wide range of frequency.

The paper is organized as follows. In the first section, the correspondence between quantum EIT phenomenon and its classical two-oscillator model is outlined. In the second section, the description of the Doppler shift in an EIT medium is presented, with emphasis on the sensitivity of the shift to the system parameters. Finally, the Finite Difference Time Domain method is used to confirm the theoretical findings.

\section{Electromagnetically induced transparency in Lambda system and its classical analog}
\begin{figure}[ht!]
\includegraphics[scale=0.24]{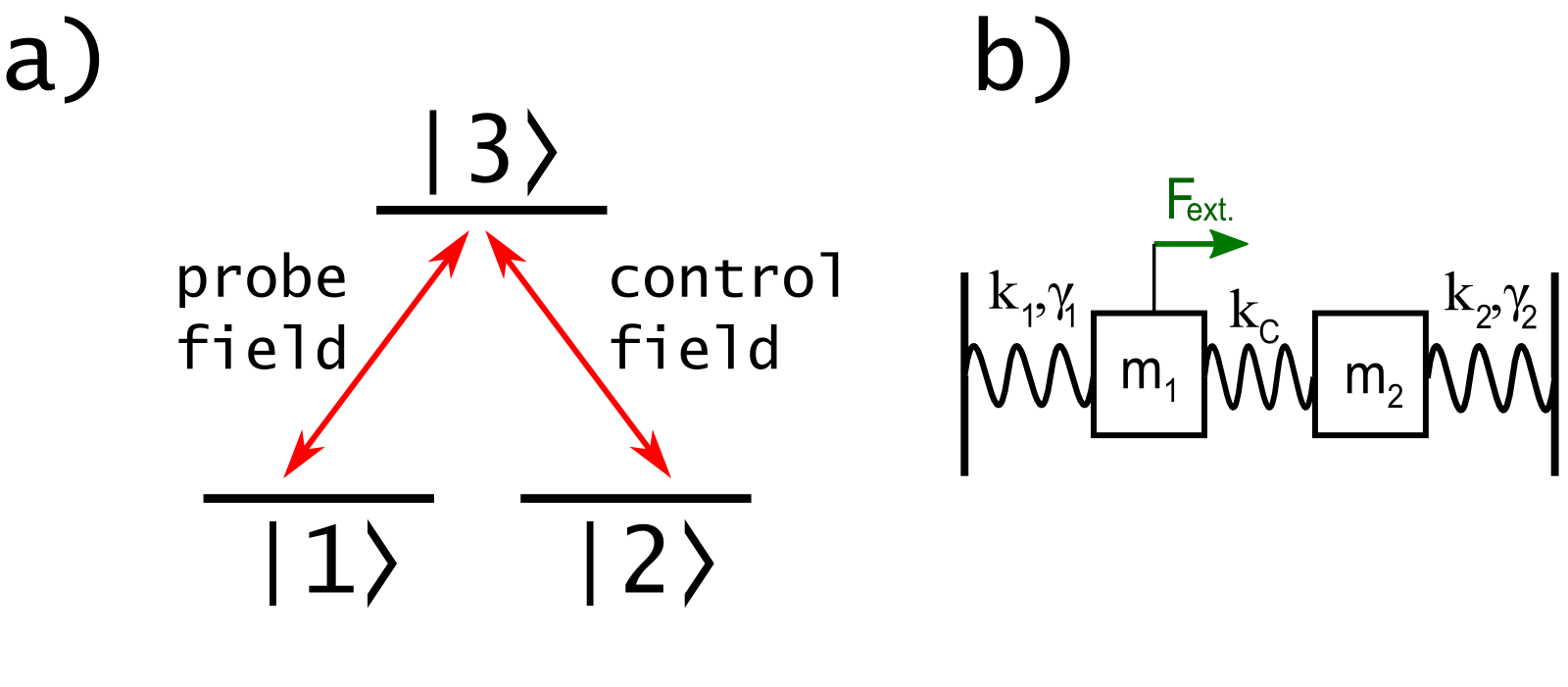}
\caption{a) Schematic representation of a typical EIT system in lambda configuration. b) Mechanical analogue of the system.}\label{fig:1}
\end{figure}

  The electromagnetically induced transparency (EIT) is a quantum interference phenomenon where applied electromagnetic control field generates a narrow frequency region where absorption of a probe beam is suppressed. This effect is accompanied by a very steep frequency dispersion curve. It means that the medium has become transparent for the probe pulse which travels with the reduced group velocity.
 The generic atomic system where EIT is observed in a three-level atomic medium consists of atoms in so-called lambda configuration, shown on the Fig. \ref{fig:1} (a).

The  system consists of two ground levels $|1\rangle$ and $|2\rangle$ which have dipole-allowed optical transitions to the excited state $|3\rangle$ and the population of the system is initially in the ground state $|1\rangle$. The atom-light coupling scheme involves two laser fields, one probe characterized by Rabi frequency $\Omega_1$ of low intensity and a control  of much higher intensity with Rabi frequency $\Omega_2$, where both Rabi frequencies account for the strength of atom-field coupling for a given transitions, $\Omega _i=\frac{d_{i3}E_i}{\hbar}$, $i=1,2$.

 The optical properties of the medium are described by complex electric susceptibility $\chi$: its real part is related to the index of refraction of the medium and the imaginary part depicts absorption. The electric susceptibility for the $\Lambda$ system is given by \cite{Scully}
\begin{equation} \label{EIT2}
\chi(\omega)=\epsilon_r(\omega)-1=\frac{-N|d_{13}|^2}{\hbar \epsilon_0}\frac{1}{\omega - \omega_{1} + i\Gamma_{31} - \frac{\Omega_2^2}{\omega - \omega_{2} + i\Gamma_{12}}},
\end{equation}
where $\omega_1$ and $\omega_2$ are the frequencies associated with $\left|1\right>\rightarrow\left|3\right>$ and $\left|2\right>\rightarrow\left|3\right>$ transitions, $\Gamma_{31}$, $\Gamma_{12}$ are the relaxation rates and $|d_{13}|$ is a mean value of the dipole moment operator of the $\left|1\right>\rightarrow\left|3\right>$ transition.

 It is known that  EIT-like phenomena can in principle be observed in classical systems due to the fact that some aspect of the atom-field interaction can be described by the classical theory of optical dispersion \cite{Souza}, and no quantum mechanical states are necessary to observe EIT in metamaterials.
 Recently, it has been shown that such system supports a dark state leading to phenomena similar to EIT, can be represented by a wide variety of classical analogues: plasmonic structure \cite{Zhang}, metal-superconductor hybrid metamaterial \cite{Kurter}, coupled microresonators \cite{Smith} or cut wire-pairs \cite{Tamayama}.
 The simplest and  the most intuitive model of such a system consists of two harmonic oscillators coupled by a spring, shown on the Fig. \ref{fig:1} (b) \cite{Souza}. One mass characterized by dissipation factor $\gamma_1$ is acted upon by an external force, while the other mass with damping constant $\gamma_2$ is not driven directly. If $\gamma_2<<\gamma_1$ such a system reproduces an EIT-like absorption dip and steep dispersion in its power spectrum due to destructive interference between the normal modes at the resonance frequency \cite{Tassin2}. In analogy to the generic lambda system each atomic transition is represented by harmonic oscillator which loses energy by some mechanical friction mechanism and the transition frequencies $\omega_1$ and $\omega_2$ correspond to natural frequencies of both oscillators. The control and the probe fields are identified by coupling spring and by harmonic external force acting on mass $m_1$, respectively. The detailed  analysis of analogy between lambda system and two mechanical coupled harmonic oscillators was recently presented by Souza \emph{et al} \cite{Souza}.
 In EIT metamaterials, the displacements of the masses in the described model can be translated into displacements of charges, generating polarization vectors $\vec P_1$ and $\vec P_2$, corresponding to the $|1\rangle \rightarrow |3\rangle$ and $|2\rangle \rightarrow |3\rangle$ transitions \cite{Griffiths}. These polarizations are described by equations
\begin{eqnarray} \label{ClassEit1}
\ddot{P_1} + \gamma_1 \dot{P_1} + \omega_{1}^2 P_1  - f_2 P_2 &=& \epsilon_0 f_1E, \nonumber\\
\ddot{P_2} + \gamma_2 \dot{P_2} + \omega_{2}^2 P_2  - f_2 P_1 &=& 0,
\end{eqnarray}
where $\gamma_1$, $\gamma_2$ are the damping constants, $f_1$ is the coupling between the system and external field, and $f_2$ is the coupling between oscillators. For harmonic external field $E = E_0 e^{-i\omega t}$, a steady state solution in the form $P_1 = P_{01}e^{-i \omega t}$, $P_2 = P_{02}e^{-i \omega t}$ yields the following dispersion relation
\begin{equation} \label{EIT3}
\chi_e(\omega)=\frac{P_1}{\epsilon_0 E}=\frac{f_1}{\omega_{1}^2 - \omega^2 - i\gamma_1 - \frac{f_2^2}{\omega_{2}^2 - \omega^2 - i\gamma_2}}.
\end{equation}
In  resonance, when $\omega_1 \approx \omega_2 \approx \omega$, the above formula reduces to the form similar to Eq. \ref{EIT2}, where $\Gamma_{31} = \frac{\gamma_1}{2}$, $\Gamma_{12} = \frac{\gamma_2}{2}$, $\frac{N|d_{13}|^2}{\hbar \epsilon_0} = \frac{f_1}{2\omega}$, $\Omega_2 = \frac{f_2}{2\omega}$. Due to this analogy one can see that the optical control of the EIT system realized by changes of the strength of the coupling field can be obtained by modification of the constant $f_2$.
Due to the dependence on electric susceptibility of the refractive index $n=\sqrt{1+Re\chi}$ and the group index $n_g = n + \omega\frac{\partial n}{\partial \omega}$, both the group velocity and the phase velocity of the probe fields
\begin{equation} \label{Vg}
V_g(\omega) = \frac{c}{n_g(\omega)} =\frac{c}{n + \omega\frac{\partial n}{\partial \omega}},\qquad V_p(\omega) = \frac{c}{n(\omega)}
\end{equation}
are influenced by $\chi$. The susceptibility is, in turn, closely related to intensity of coupling  coefficient $f_2$ (the Rabi frequency of a control field $\Omega_2$ ). Therefore, the control field can be used to manipulate the conditions for  the probe field propagation, i.e. manipulating of the macroscopic parameter $f_2$ results in tunability of EIT. An example of a classical system where such control is provided are the so-called tunable metamaterials, demonstrated by Gu \emph{et al} \cite{Gu}, where by incorporating Si islands into the metamaterial unit cell, it was possible to demonstrate EIT effect in planar metamaterials functioning at terahertz regime. Electrically controlled systems are also realized by the introduction of variable capacitance element into the metamaterial structure \cite{Feng}.

\section{Doppler shifts ~in~EIT medium}

\subsection{ The model}

Here we consider the radiating source moving inside EIT medium.
It has been shown in \cite{Ziemkiewicz} that the material dispersion plays an important role in the mechanism of the Doppler shift of frequency for the source moving inside the medium, leading to the so-called complex Doppler effect where monochromatic source generates wave modes of multiple frequencies.
When a monochromatic source having frequency $\omega_0$ moves with velocity $V$ relative to the medium, the shifted frequency $\omega$ in a stationary detector can be obtained from the relation \cite{Ziemkiewicz2}
\begin{equation}\label{r_Dopler}
\omega_0 \sqrt{1 - \beta^2} = \omega - \vec k \cdot \vec V,
\end{equation}
where $\beta = V/c$ and $\vec k$ is the wave vector in the reference frame of the detector. For any given source velocity $V$, the left hand side of Eq. \ref{r_Dopler} is a constant, while the right hand side is a function of frequency. In the case when the wave is propagating in free space without dispersion, it is a linear function $f(\omega) = \omega(1 - \beta)$ which reaches the value $\omega_0\sqrt{1 - \beta^2}$ once, generating single, Doppler shifted frequency $\omega=\omega_0\sqrt{1+\beta}$. On the other hand, in dispersive media the relation $k(\omega)$ becomes nontrivial, and the Eq. \ref{r_Dopler} might have multiple solutions.
EIT even in relatively simple systems leads to the situation  where dispersion relation of the medium is both complicated and optically controlled, possibly leading to many interesting effects connected with the generation and propagation of the waves emitted by a moving source.

\begin{figure}[ht!]
\begin{minipage}[b]{.45\linewidth}
\includegraphics[scale=0.45]{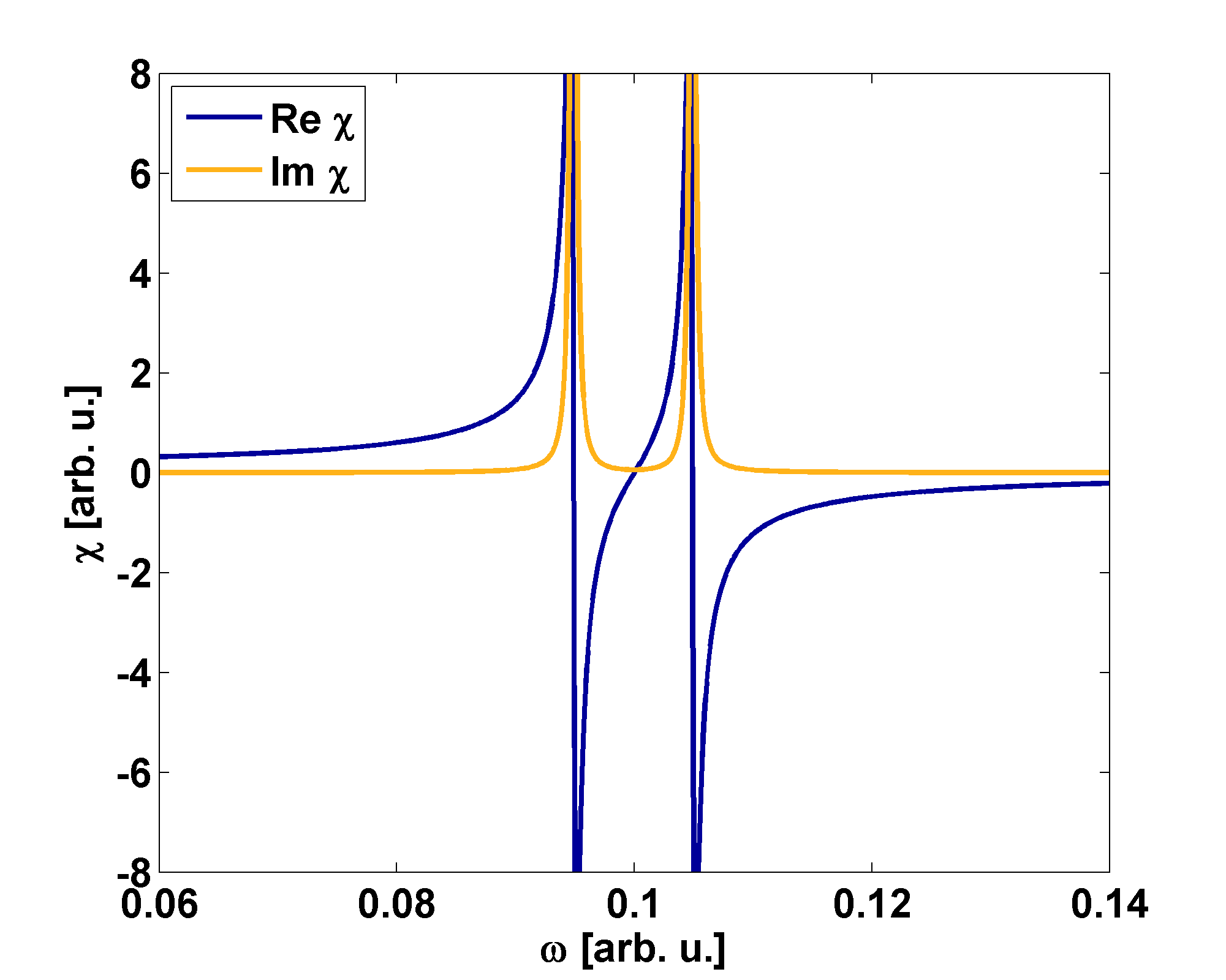}
\caption{Dispersion relation of the EIT medium   with the parameters $ f_1=0.002~\gamma_1=0.0006~\omega_{1}=0.1~f_2=0.001~\gamma_2=0.0003~\omega_{2}=0.1$. }\label{fig:2a}
\end{minipage}
\begin{minipage}[b]{.45\linewidth}
\includegraphics[scale=0.45]{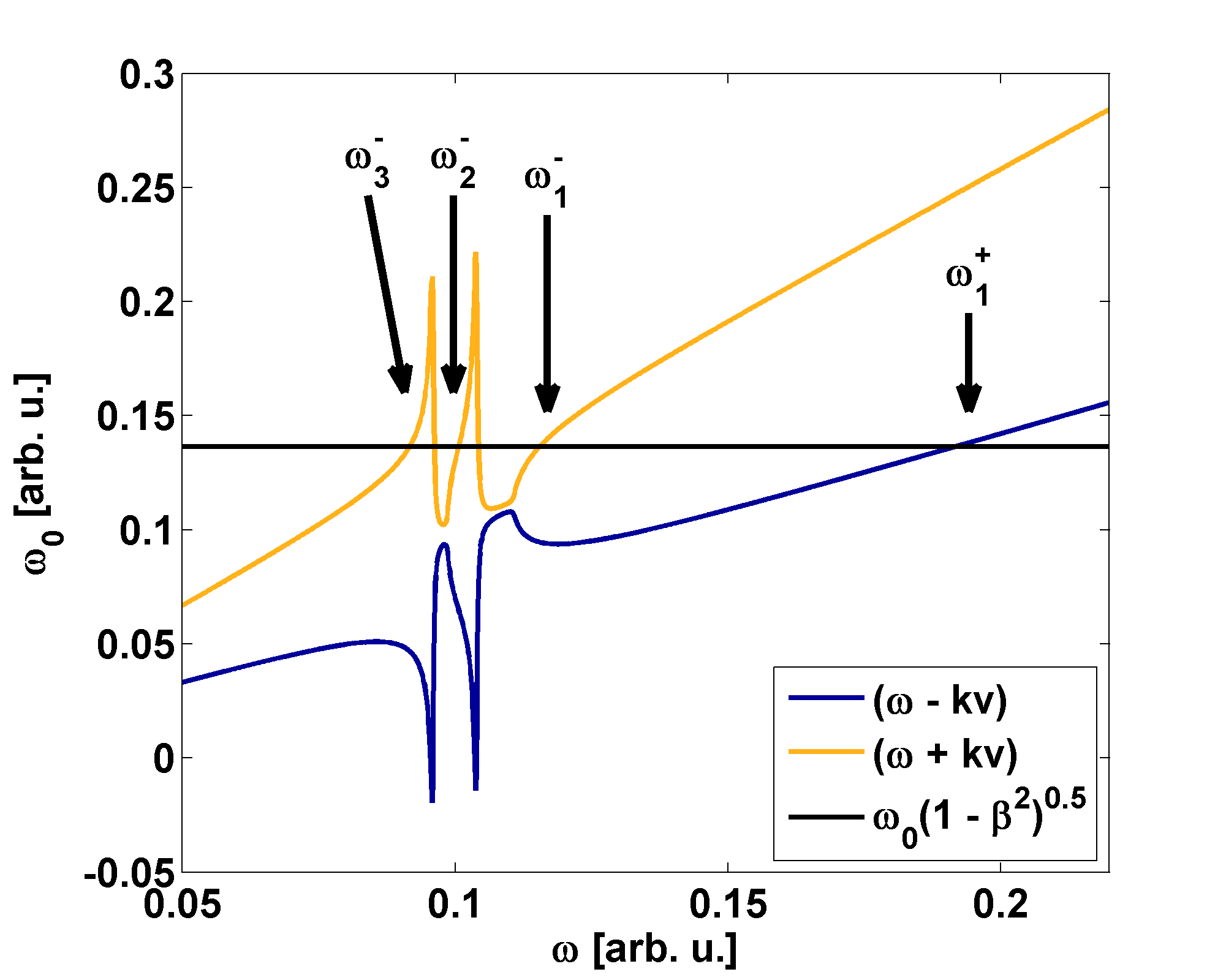}
\caption{Plot of the left and right hand side of Eq. \ref{r_Dopler} in a one dimensional case.\newline\newline}\label{fig:2b}
\end{minipage}
\end{figure}

Lets consider a medium of susceptibility given by Eq. \ref{EIT3}.
The dispersion relation $\chi(\omega)$ for such medium is shown on the Fig. \ref{fig:2a} and exhibits a narrow transparency window. The parameters were set to obtain a typical dispersion curve for lambda system under EIT condition. A well-known feature of EIT is that the width of the transparency window is proportional to the square of the control field amplitude and the normal dispersion inside the window is responsible for reduction of the group velocity. In order to describe the method we use for our analysis we first consider the case of a source traveling through EIT media with the speed $V=0.3 c$ emitting the signal of frequency $\omega_0 = 0.15$ which is outside  the transparency window, in the region where $Re(\chi(\omega_0))\approx Im(\chi(\omega_0)) \approx 0$ (see Fig. \ref{fig:2a}). In a simplest, one-dimensional case, one can consider only waves emitted in the forward direction ($\vec k \cdot \vec V = kV$) and backward direction ($\vec k \cdot \vec V = -kV$). For these two cases, one can find the Doppler frequency modes at a given speed $V$ by comparing the left and the right hand side of Eq. \ref{r_Dopler} on a plot, as shown on the Fig. \ref{fig:2b}. For any given $V$ and $\omega_0$, the frequency modes $\omega$ can be read from the $x$ axis at the points where the constant value of $\omega_0\sqrt{1-\beta^2}$ is equal to $f^\pm(\omega)=\omega \mp kV$. Another, general approach of solving Eq. \ref{r_Dopler} is presented in Appendix.

The usual, upshifted wave emitted in the direction of source motion is denoted by $\omega_1^+$. One can see that around this frequency, the function $f^+(\omega)=\omega - kV$ is almost linear which is consistent with the negligible dispersion at this point. The $\omega_1^-$ mode describes
 the standard, downshifted wave emitted in the backward direction. However, one can see that due to the characteristic dispersion relation of the EIT medium, the function $f^-(\omega)=\omega + kV$ crosses the value of $\omega_0\sqrt{1 - \beta^2}$ four more times. Two of those solutions, denoted $\omega_2^-$ and $\omega_3^-$, are located in the region of normal dispersion ($\frac{\partial k}{\partial \omega}>0$), where absorption is small. The mode $\omega_2^- \approx 0.1$ is located inside the transparency window and the $\omega_3^- \approx 0.09$ is below its edge. By examining the plot of the function $f^{\pm}(\omega)$ on the Fig. \ref{fig:2b}, several characteristic features can be deduced. Since the x and y axis of the plot correspond to the shifted frequency and the source frequency respectively, one can use the derivative of the function $f^{\pm}(\omega)$ to obtain the sensitivity of the Doppler shift to the source frequency
\begin{equation}
\frac{\partial \omega}{\partial \omega_0} = \left(\frac{\partial f^\pm}{\partial \omega}\right)^{-1}.
\end{equation}
The function $f^\pm=\omega \mp kV$ is closely related to the material dispersion by the term $kV$. In the frequency regions where dispersion curve is very steep, such as inside the transparency window, the Doppler shifted frequencies will be insensitive to the changes of the source frequency. This can be also confirmed analytically; by taking derivative of Eq. \ref{r_Dopler}, for the one-dimensional case one obtains
\begin{equation} \label{dwdw0}
\frac{\partial \omega^\pm}{\partial \omega_0} = \frac{\sqrt{1-\beta^2}}{1 \mp \frac{V}{V_g}}.
\end{equation}
Therefore, the small response to the changes of the source frequency is realized when the group velocity is small compared to the source velocity. This is easily realized inside the transparency window. In the described example, $V_g(\omega_2^-) << V=0.3~c$ and one can see on the Fig. \ref{fig:2b} that even significant changes of $\omega_0$ push the value of $\omega_2^-$  only very little. On the other hand, the modes $\omega_1^\pm$ are located in the area where dispersion is negligible and the function $f^\pm(\omega)$ is linear. One can conclude that in the absence of material dispersion, there would be only two solutions $\omega_1^\pm$, consistent with the Doppler shift in vacuum.

\subsection{ Numerical simulations and discussion  for one dimensional system}
In order to solve this problem the Finite Difference Time Domain method has been used to simulate a system consisting of a point radiation source moving along $\hat{x}$ axis. As in our previous work \cite{Ziemkiewicz}, the source was modeled as a spatially narrow, Gaussian shaped area of electric current. The response of the medium to the electromagnetic field is calculated with Auxiliary Differential Equations (ADE) method, where two polarizations are given by Eq. \ref{ClassEit1}. The parameters (from Fig.2) were set to obtain a typical dispersion curve characteristic to EIT phenomenon and to minimize the effect of numerical dispersion \cite{Taflove}. The medium is considered isotropic and magnetically inactive ($\mu=1$), with positive refraction index $1 < n < 4.5$ and maximum group index $n_g \approx 200$. As reported in \cite{Jahromi}, such values can be obtained in metamaterials.

\begin{figure}[ht!]
\begin{minipage}[b]{.45\linewidth}
\includegraphics[scale=0.45]{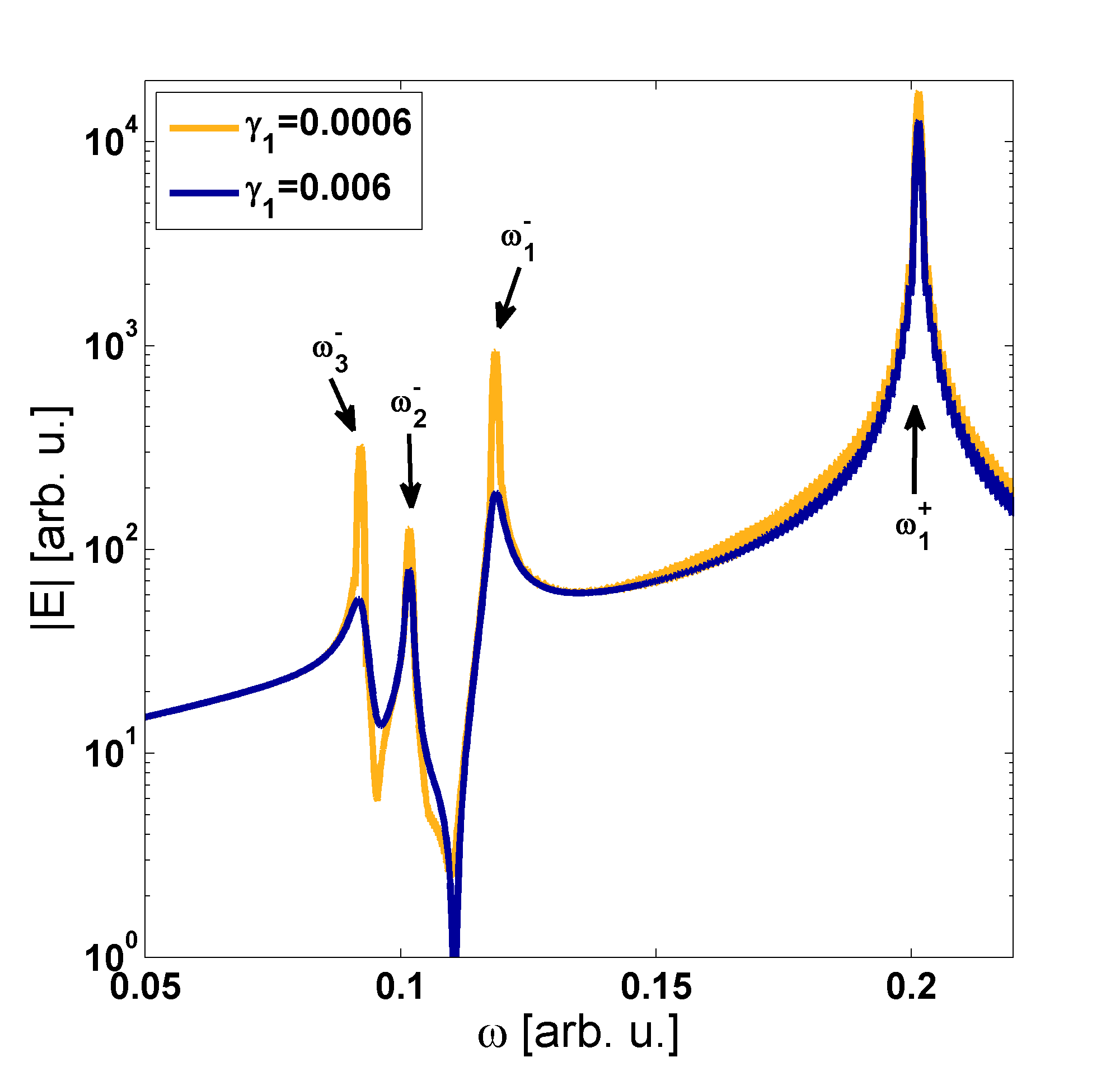}
\caption{Radiation spectrum of a source moving with velocity $V=0.3~c$ for different values of damping constant  $\gamma_1$.\newline}\label{fig:3a}
\end{minipage}
\begin{minipage}[b]{.45\linewidth}
\includegraphics[scale=0.45]{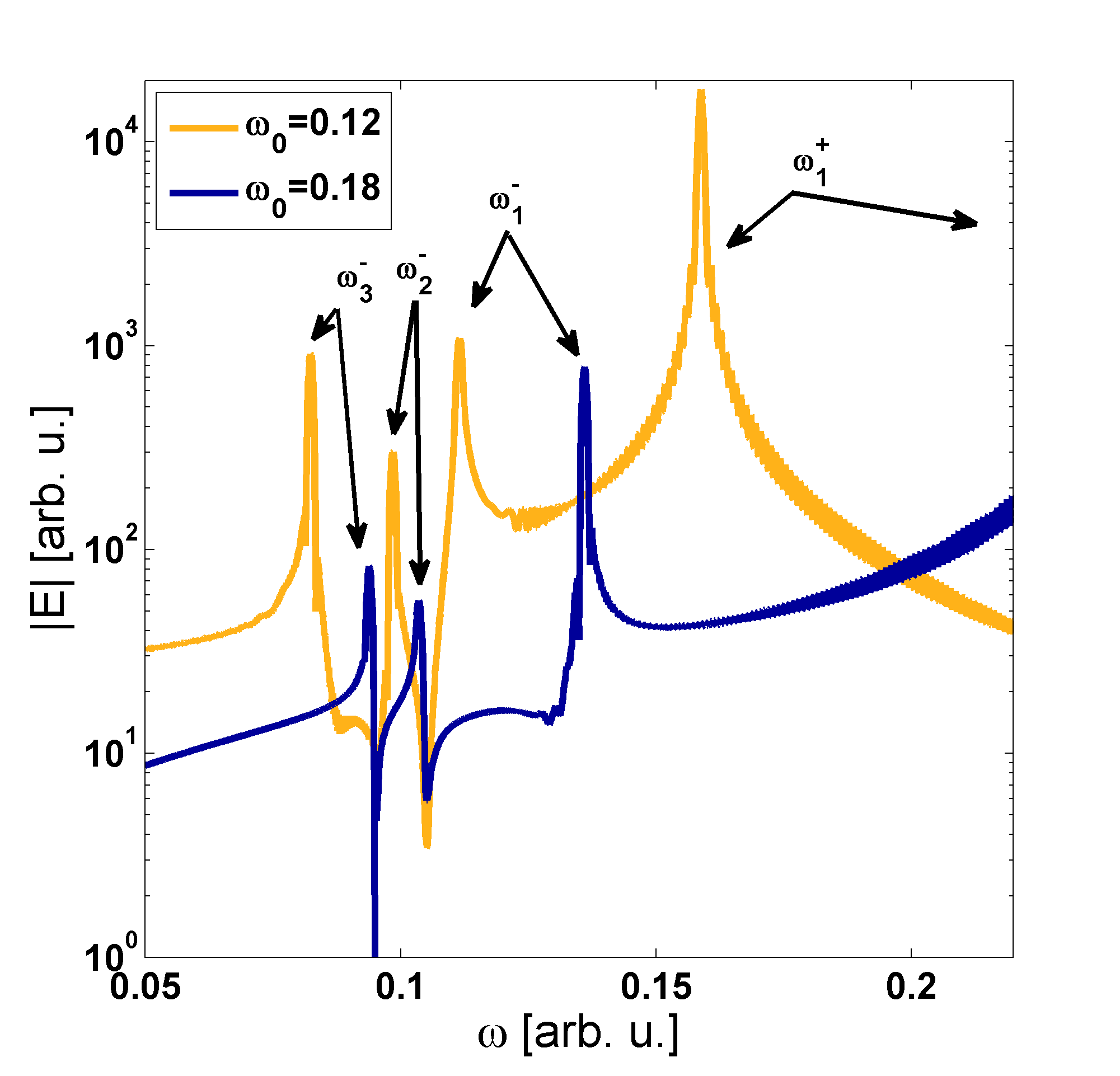}
\caption{Radiation spectrum  as a function of the source frequency for the source velocity $V=0.3c$ and for two different source frequency $\omega_0$.}\label{fig:3b}
\end{minipage}
\end{figure}

For the case of $V=0.3~c$ presented on the Fig. \ref{fig:2b}, the frequency spectrum of the emitted field was obtained by a fast Fourier transform and the results are shown on the Fig. \ref{fig:3a}. For comparison, two values of damping constant $\gamma_1$ were used. There is single, strong, upshifted mode $\omega_1^+$ emitted in the forward direction. It is located far beyond of the transparency window, in the region where medium becomes almost dispersionless (i.e. $Re \chi \approx 0$). Therefore, it is almost unaffected by the changes of the medium model parameters. The energy emitted in the backward direction is divided between three modes $\omega_1^-$, $\omega_2^-$, $\omega_3^-$ with comparable amplitudes. The mode $\omega_2^-$ located inside the transparency window is easily detectable due to low or even zero absorption. It is relatively insensitive for increasing damping  $\gamma_1$, which also determines the shape and  transparency windows width. One can also observe significant dips in the obtained spectrum around $\omega_2^-$ which correspond to frames of transparency window where there are maxima of absorption (Fig. \ref{fig:2a}).
To illustrate the dependence of the Doppler modes on the source frequency, the simulation has been performed for $\omega_0=0.12$ and $\omega_0=0.18$ and the results are shown on the Fig. \ref{fig:3b}.
One can see that with changes of the source frequency, the general form of the spectrum is preserved. As expected, the usual Doppler mode $\omega_1^+$ is significantly shifted. On the other hand, the frequency $\omega_2^-$ located inside the transparency window is almost the same for the two presented cases, confirming previous findings. The Doppler frequencies match the solution obtained from Fig. \ref{fig:2b} by setting $\omega_0=0.12$ and $\omega_0=0.18$, respectively.

\begin{figure}[ht!]
\begin{minipage}[b]{.38\linewidth}
\includegraphics[scale=0.4]{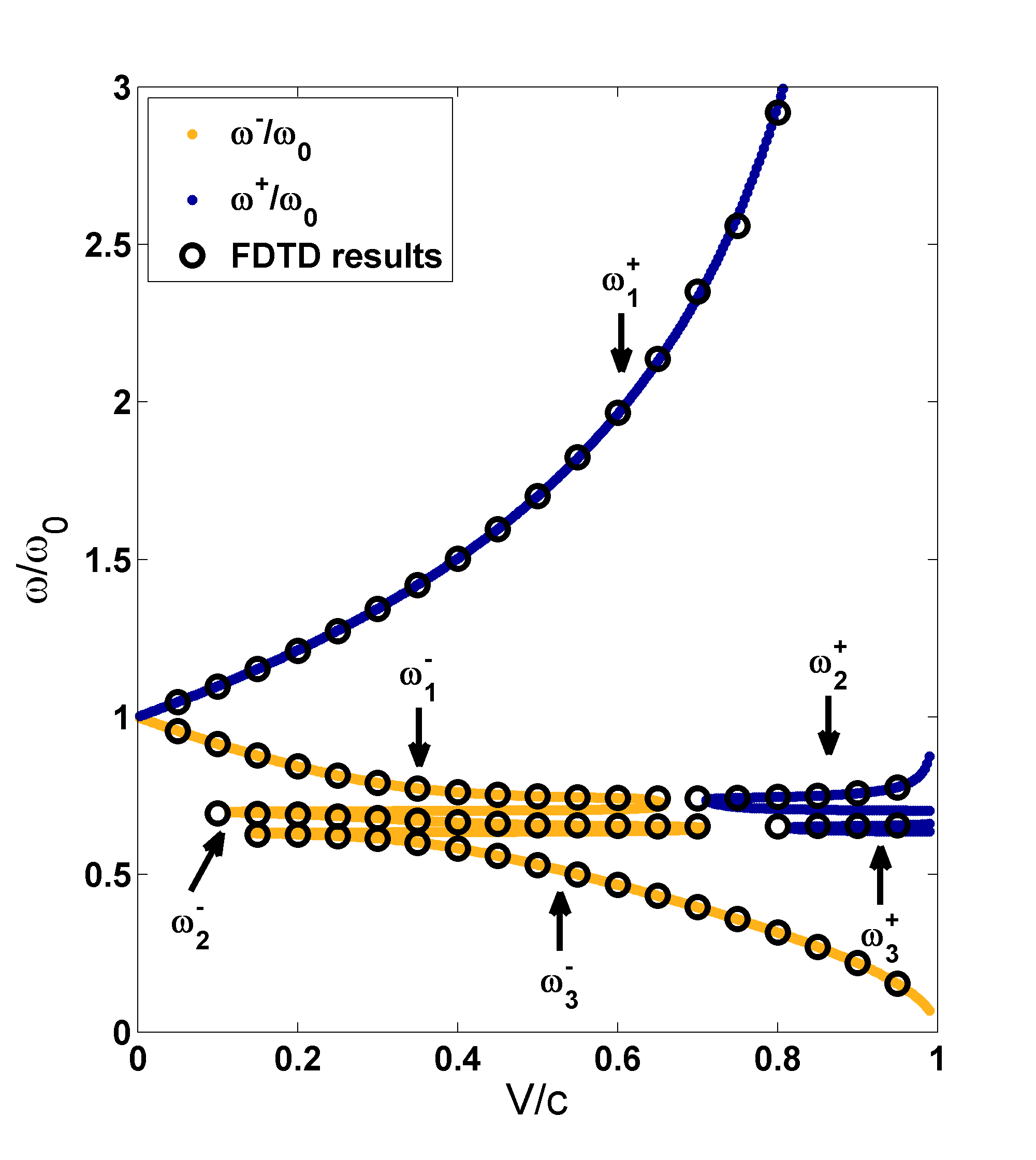}
\caption{The frequency of the Doppler modes as a function of the source speed for $\omega_0=0.15$.}\label{fig:4a}
\end{minipage}
\begin{minipage}[b]{.52\linewidth}
\includegraphics[scale=0.35]{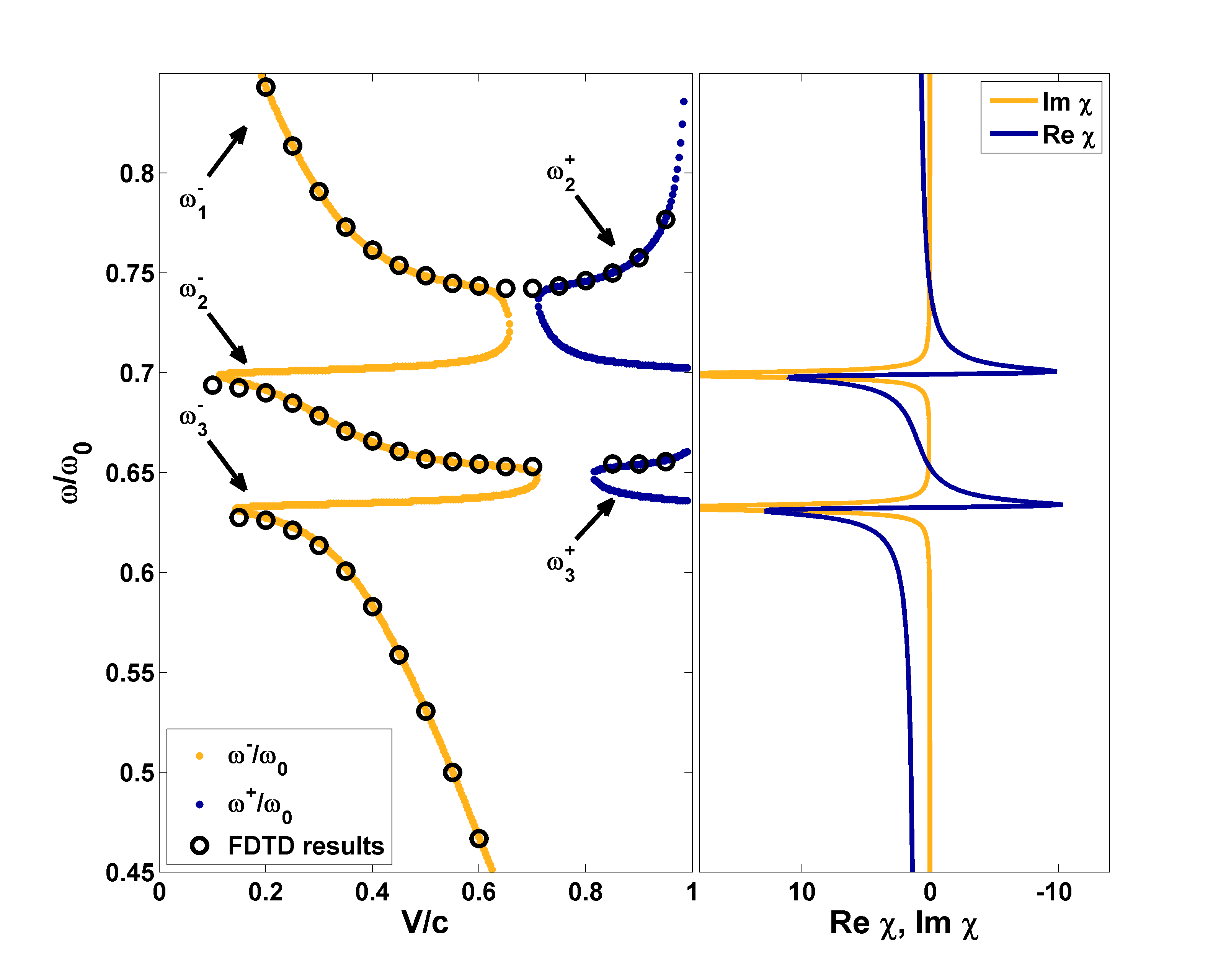}
\caption{Doppler modes on the edge and inside the transparency window. Material dispersion relation is added for reference.}\label{fig:4b}
\end{minipage}
\end{figure}

The Doppler shift relation in Eq. \ref{r_Dopler} has been solved for $\omega_0=0.15$ and a range of source velocities and the result is presented on the Fig. \ref{fig:4a}. When $V<0.1~c$, there is only single upshifted and downshifted wave. However, as the source speed increases, the downshifted wave enters the transparency window at $\omega=0.1\approx0.67~\omega_0$. Then, in a considerable range of source speeds $0.1~c < V < 0.7~c$, there are three modes emitted in the backward direction. One can see that the modes located near the window, in the regions of steep dispersion, have almost constant frequency. For the high source speed $V>0.75~c$, there are two additional forward emitted modes $\omega_2^+$ and $\omega_3^+$.
  The behavior of the modes located near and inside the window is of special interest and has been shown in detail on the Fig. \ref{fig:4b}. The dispersion relation of the medium was added for reference.
One can see that only the modes located in the regions of normal dispersion were detected in FDTD simulation.
Moreover, as the dispersion curve $Re \chi(\omega)$ becomes steeper, the variation of the frequency with the source speed decreases. For example, the mode $\omega_2^-$ located inside the window has almost constant frequency in a significant range of source speed. This can be confirmed analytically. The left and right hand side of Eq. \ref{r_Dopler} can be written as
\begin{eqnarray}
f(\beta)&=&\omega_0\sqrt{1 - \beta^2}\nonumber\\
g(\beta,\omega)&=&\omega - \omega n(\omega)\beta
\end{eqnarray}
then, from the exact differential at the point $\omega_1$, $\beta_1$, one obtains
\begin{eqnarray}
df(\beta_1)&=&\frac{-\omega_0 \beta_1}{\sqrt{1-\beta_1^2}}d\beta\nonumber\\
dg(\beta_1,\omega_1)&=&(1 - n\beta_1 - \omega_1\frac{\partial n}{\partial \omega}\beta_1)d\omega - \omega_1 n d\beta,
\end{eqnarray}
and finally
\begin{equation}\label{R1}
\frac{\partial \omega}{\partial \beta}\vert_{\omega_1,\beta_1} = \frac{\frac{-\omega_0 \beta_1}{\sqrt{1-\beta_1^2}} + \omega_1 n}{1 - \frac{c \beta_1}{V_g}}
\end{equation}
As in the case of Eq. \ref{dwdw0}, for $V_g<<V$, the derivative vanishes. On the other hand, for $V_g = V$, one obtains $\frac{\partial \omega}{\partial \beta} \rightarrow \infty$. This means that the function $\omega(\beta)$ becomes vertical. Several such singular points can be seen on the Fig. \ref{fig:4b}, where for some velocity $V$, two modes located in the regions of normal and anomalous dispersion meet, forming a loop (as for example for the backward mode  $\omega_1^-$ at $V=0.65~c$, see Fig. 7). The source velocity range between such loops represents a cutoff, which separates  the propagating and nonpropagating waves inside transparency window. It means that for some range of the source velocities the radiation  of frequency well-fitting transparency window is not emitted. To sum up, in the complex Doppler effect, the cutoff velocities correspond to the points where the group velocity of the generated frequency mode is equal to the source velocity. Such a case was observed in a simple Drude model medium in \cite{Ziemkiewicz}.

In considered case, for smaller than cutoff velocity, there are three backward frequency modes (one of them $\omega_2^-$  is located inside  the transparency window) and for higher source velocities situation becomes opposite and one backward mode $\omega_3^-$ and three forward modes  ($\omega_1^+,\omega_2^+$ and $\omega_3^+$, the last is located inside the window) appear.
 To conclude, the wave modes on the edge of transparency window and inside it are generated in a wide range of source speeds and frequencies and show little variance with these parameters and cutoffs concern only  frequency modes inside the transparency window. These properties are strongly connected with the dispersion relation of the medium.

\begin{figure}[ht!]
\begin{minipage}[b]{.38\linewidth}
\includegraphics[scale=0.4]{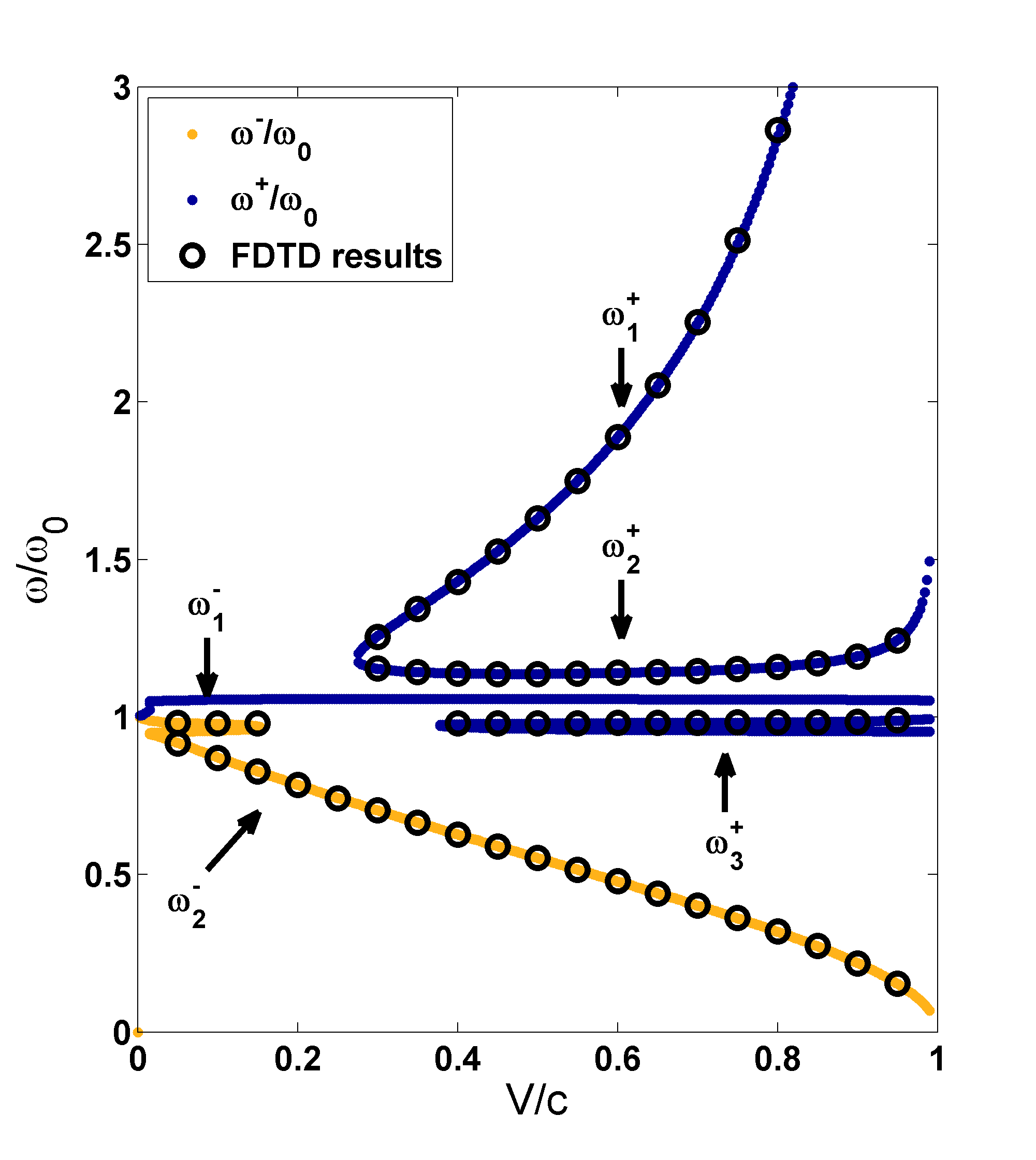}
\caption{The frequency of the Doppler modes as a function of speed for $\omega_0=0.1$.}\label{fig:5a}
\end{minipage}
\begin{minipage}[b]{.52\linewidth}
\includegraphics[scale=0.35]{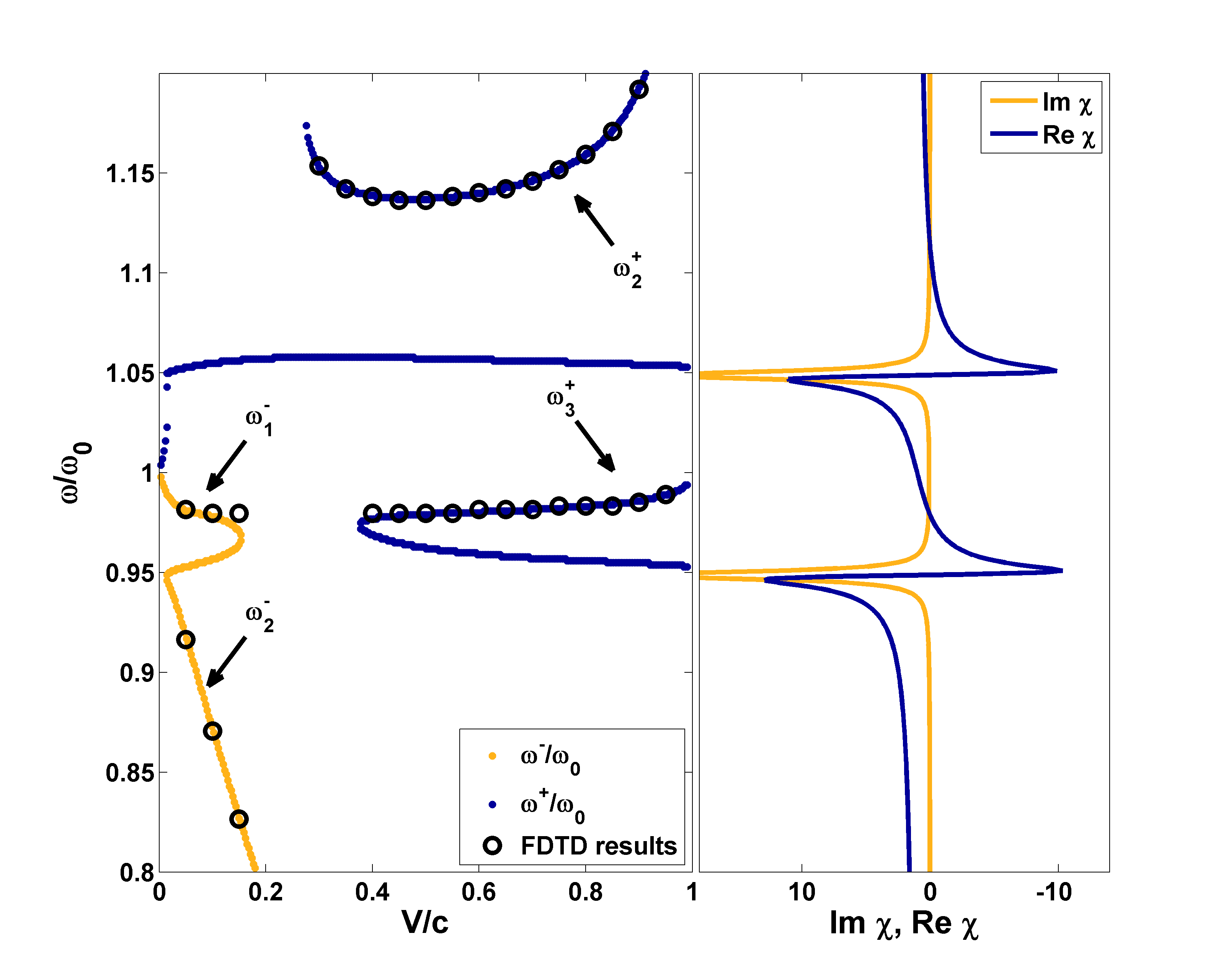}
\caption{Doppler modes on the edge and inside the transparency window. Material dispersion relation is added for reference.}\label{fig:5b}
\end{minipage}
\end{figure}

\begin{figure}[ht!]
\begin{minipage}[b]{.45\linewidth}
\includegraphics[scale=0.45]{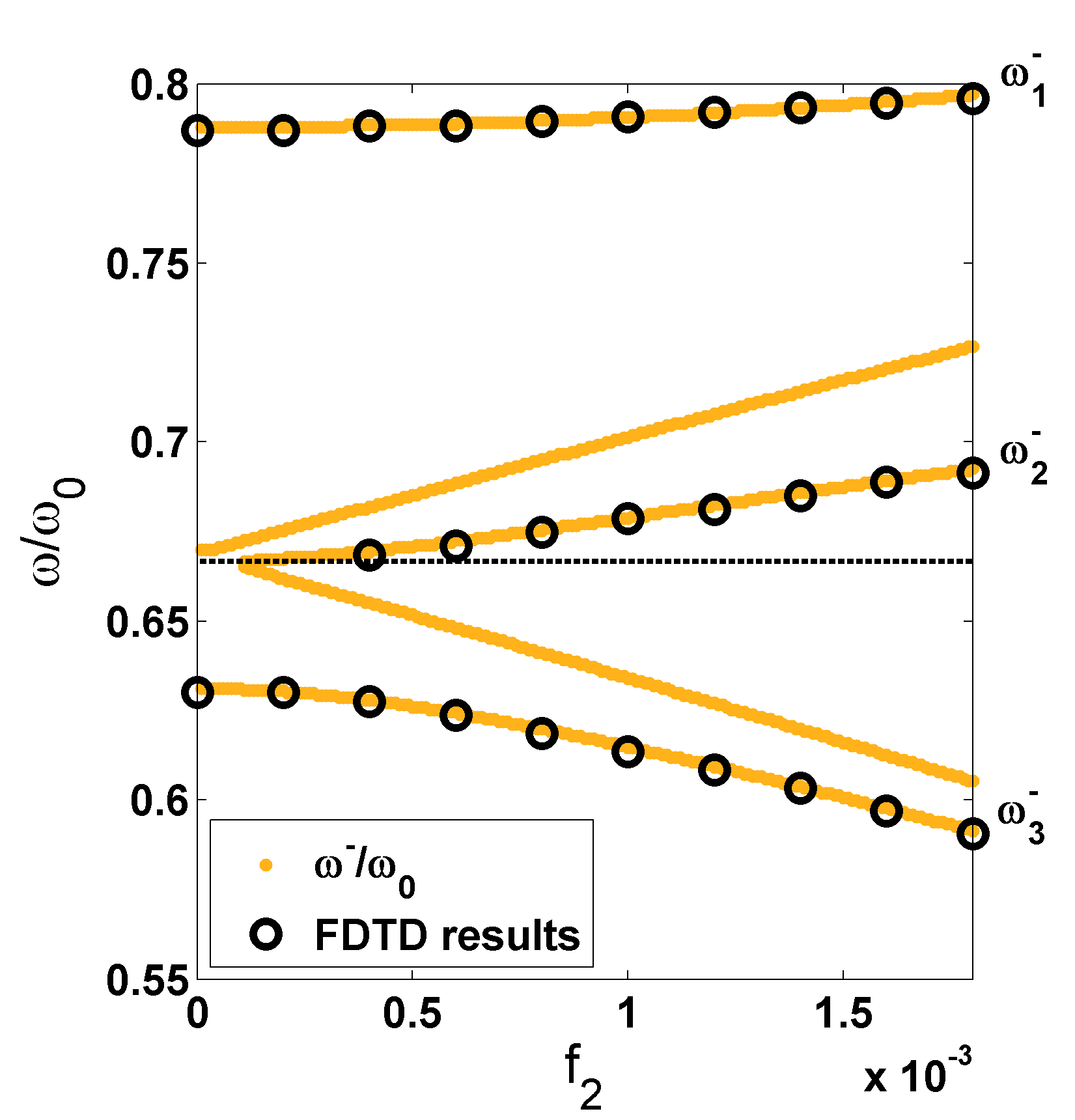}
\caption{The dependence of the Doppler shifts on the width of the transparency window controlled by coupling $f_2$ for $\omega_0=0.15$.}\label{fig:6a}
\end{minipage}
\begin{minipage}[b]{.45\linewidth}
\includegraphics[scale=0.45]{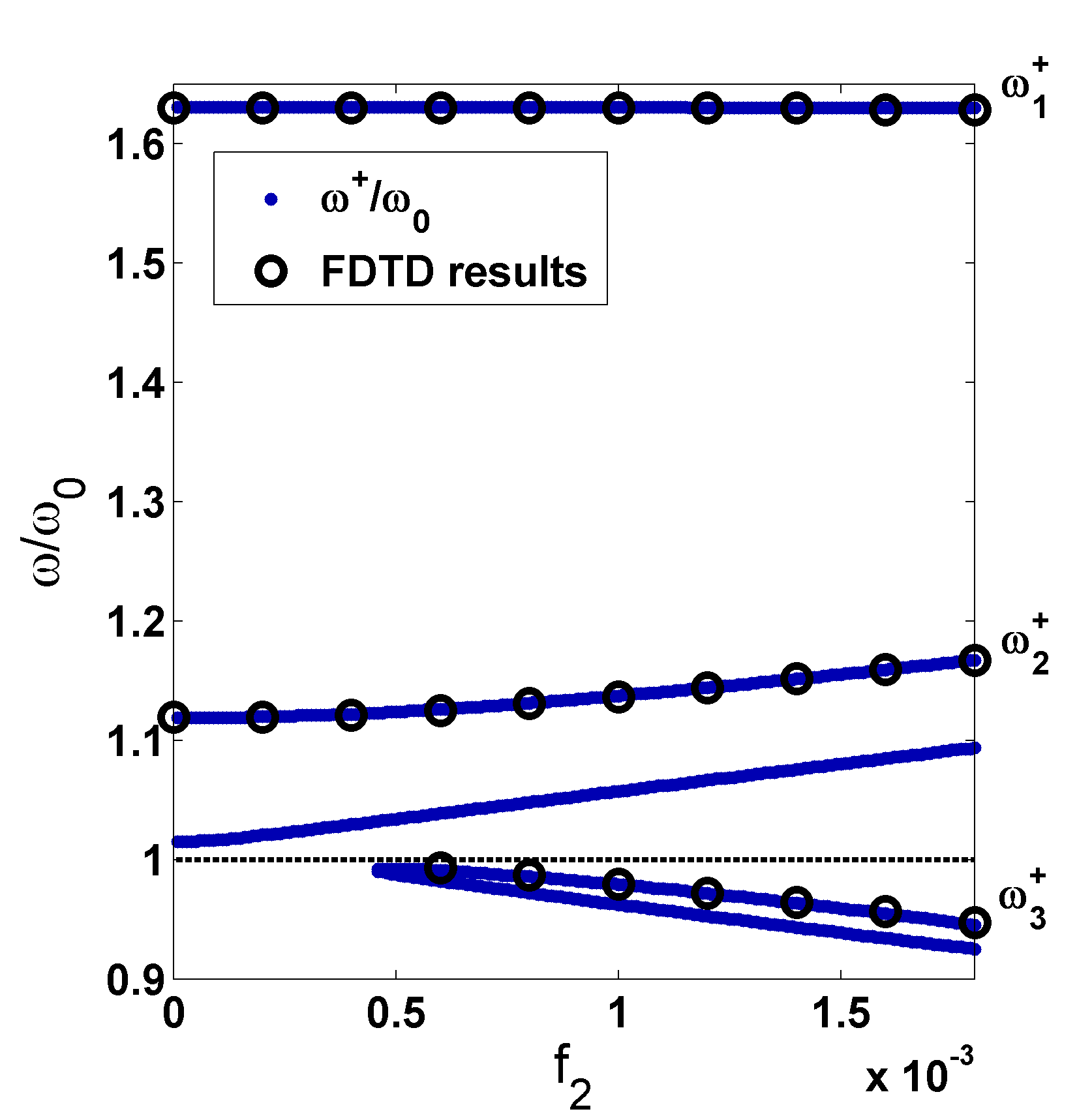}
\caption{The dependence of the Doppler shifts on the width of the transparency window controlled by coupling $f_2$ for $\omega_0=0.1$.}\label{fig:6b}
\end{minipage}
\end{figure}

 The important case is the situation where the frequency of the source is located inside the transparency window. The results of the simulation for $\omega_0=0.1$ are shown on the Fig. \ref{fig:5a} and Fig. \ref{fig:5b}. The forward emitted wave $\omega_3^+$ fits the transparency window and shows little variation with velocity. The usual Doppler modes $\omega_1^+$ and $\omega_1^-$ are located outside of the window and the relation $\omega^\pm_1(V)$ is almost linear due to negligible dispersion. For the lower cutoff velocity  $V<0.25~c$, no waves emitted in the forward direction are present.
In the case of the source frequency fitting the transparency window there is a range of velocities for which any mode appears in a region of transparency and the cutoff between the loops is wider than in the previously discussed situation. For source velocities smaller than cutoff one can see only two backward emitted modes while for velocities above the second cutoff there are three forward modes ($\omega_1^+,\omega_2^+$ and $\omega_3^+$) and one backward mode $\omega_2^-$.
\newpage
The characteristic feature  of the EIT system is the possibility of  optical control of the transparency window width which  increases with the coupling strength ($\sim\Omega^2)$.  The  modification of the dispersion relation by variation of the coupling strength opens the way to  change  Doppler shift on demand. The influence of the window width on the generated Doppler modes for $\omega_0=0.15$ and $V=0.3~c$ is shown on the Fig. \ref{fig:6a}. One can see that for the wider window the separation between downshifted modes $\omega_2^-$ and $\omega_3^-$ increases. As expected, the mode $\omega_2^-$ located inside the window becomes detectable only for sufficiently strong coupling. On the other hand, the mode $\omega_1^-$ which is located outside the window, is unaffected by the coupling strength. These findings are also confirmed for the case of the source frequency located inside the window, as is shown on the Fig. \ref{fig:6b}.
\section{Conclusions}

 We have studied the frequency shifts of radiating particle due to complex Doppler effect moving in EIT arbitrary medium. Our results are quite general and not restricted to any particular system, having application to any physical system with normal dispersion associated with low absorption. We considered the situation when the frequency of the source does not fit the transparency window and the case when the source emitted signal fitting the transparency window. We have shown that the sensitivity of the Doppler shift to the source frequency and velocity is strongly connected with the group velocity of the emitted wave. Due to the very low group velocity inside the transparency window, unusual Doppler modes occur in a wide velocity ranges separated by cutoffs dependent on the mode group velocity. We have also examined the behavior of radiation spectra and frequency shifts and their dependence on various external parameters such as coupling field intensity and damping parameters, which are responsible for the shape of transparency window and influence the dispersion of the medium. In particular, we have investigated the possibility to control the Doppler shift on demand. Our theoretical observations confirmed by FDTD simulations are proved analytically starting from the first principles. The performed simulations and  our theoretical findings give insight into the dynamic of the frequency changes of Doppler modes in the media under EIT conditions. Due to recent experiments  in  media with steep dispersion by Bortolozzo \emph{et al} \cite{Bortolozzo} it turned out that the Doppler effect in highly dispersive media might have many practical applications in precise measurements. Therefore our findings could be helpful in analysis of such measurements.

\section{Appendix}
\begin{figure}[ht!]
\includegraphics[scale=0.6]{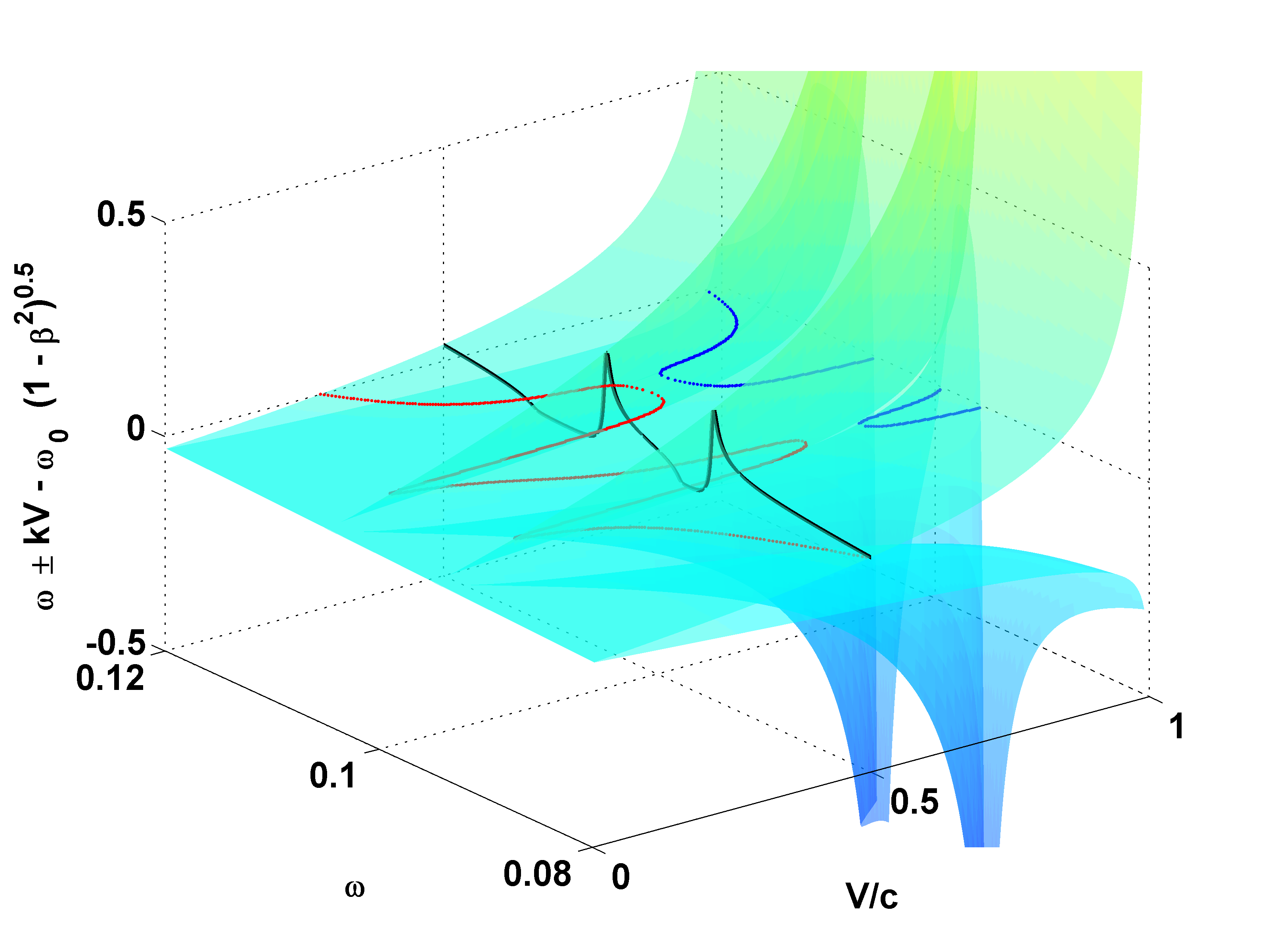}
\caption{Graphical representaton of the Doppler shift Eq. \ref{r_Dopler} in a range of source velocity $V$ and generated frequency $\omega$.}\label{fig:7}
\end{figure}

The Eq. \ref{r_Dopler} can be written in a form containing function of two variables
\begin{equation}
f^\pm(\omega,\beta) = \omega \mp kV - \omega_0 \sqrt{1 - \beta^2} = 0.
\end{equation}
The function $f$ can be plotted as a surface, where the source velocity and the frequency are on the $x$ and $y$ axis accordingly. Then, the relation $\omega(\beta)$ shown on the Fig. \ref{fig:4a} can be obtained from the intersection of the surface and the plane $z=0$, as shown on the Fig. \ref{fig:7}. By choosing a single source velocity $V$, from the intersection with $x=V=const$ plane, one obtains the functions $f^\pm(\omega)$ shown in the Fig. \ref{fig:2b}.


\begin{thebibliography}{99}
\bibitem{Boller}
K. J.  Boller, A. Imamoglu and S. E. Harris, \textit{Observation of electromagnetically induced transparency}, Phys. Rev. Lett. \textbf{66}, 2593 (1991).

\bibitem{Harris}
S. E. Harris, \textit{Electromagnetically Induced Transparency}, Phys. Today \textbf{50}, 36-42 (1997).

\bibitem{Fleisch}
M. Fleischhauer, A. Imamoglu, and J. P. Marangos, \textit{Electromagnetically induced transparency: Optics in coherent media},
Rev. Mod. Phys. \textbf{77}, 633 (2005).

\bibitem{LiuEit}
C. Liu, Z. Dutton, C. H. Behroozi, and L. V. Hau, \textit{Observation of coherent optical information storage in an atomic medium using halted light pulses}, Nature \textbf{409}, 490-493 (2001).

\bibitem{Fleisch2}
M. Fleischhauer and M. D. Lukin, \textit{Quantum memory for photons: Dark-state polaritons},
Phys. Rev. A \textbf{65}, 022314 (2002).

\bibitem{Racz1}
R. Raczy\'nski, J. Zaremba, and S. Zieli\'nska-Kaniasty, \emph{Coherent processing of a light
pulse stored in a medium of four-level atoms}, Optics Comm. 217, 275-280 (2003).

\bibitem{Racz2}
A. Raczy\'nski, J. Zaremba, and S. Zieli\'nska-Kaniasty, \emph{Electromagnetically induced transparency
and storing a pair of pulses of light}, Phys. Rev. A 69, 043801-5 (2004).

\bibitem{Racz3}
A. Raczy\'nski, J. Zaremba, and S. Zieli\'nska-Kaniasty, \emph{Beam splitting and Hong-Ou-
Mandel interference for stored light}, Phys. Rev. A 74, 031810-7 (2007).

\bibitem{Racz4}
Jin-Hiu Wu, M.Artoni, G.C. La Rocca, A. Raczy\'nski, J. Zaremba, and S. Zieli\'nska-
Kaniasty, \emph{Tunable photonic metamaterials}, J. Mod. Opt. 56, 768-781 (2009).

\bibitem{Zhang}
S. Zhang, D. A. Genov, Y. Wang, M. Liu, and X. Zhang, \textit{Plasmon-Induced Transparency in Metamaterials}, Phys. Rev. Lett. 101, 047401 (2008).

\bibitem{Souza}
J. A. Souza, L. Cabral, R. R. Oliveira, and C. J. Villas-Boas, \textit{Electromagnetically-induced-transparency-related phenomena and their mechanical analogs},
Phys. Rev. A \textbf{92}, 023818 (2015).

\bibitem{Tassin}
P. Tassin, L. Zhang, R. Zhao, A. Jain, T. Koschny, and C. M. Soukoulis, \textit{Electromagnetically Induced Transparency and Absorption in Metamaterials: The Radiating Two-Oscillator Model and Experimental Confirmation}, Phys. Rev. Lett. \textbf{109}, 187401 (2012).

\bibitem{Ziemkiewicz}
D. Ziemkiewicz and S. Zieli\'nska-Raczy\'nska, \textit{Complex Doppler effect in left-handed
metamaterials}, J. Opt. Soc. Am. B \textbf{32}, 363-369 (2015).

\bibitem{Ziemkiewicz2}
D. Ziemkiewicz and S. Zieli\'nska-Raczy\'nska, \textit{Cherenkov radiation combined with the complex
Doppler effect in left-handed metamaterials}, J. Opt. Soc. Am. B \textbf{32}, 1637-1644 (2015).

\bibitem{Bortolozzo}
U. Bortolozzo, S. Residori, and J.C. Howell, \textit{Precision Doppler measurements with steep dispersion}, Optics Letters \textbf{38}, 3107 (2015).

\bibitem{Scully}
M. O. Scully, M. S. Zubairy, \textit{Quantum optics}, (Cambridge University Press, 1997).


\bibitem{Kurter}
C. Kurter, P. Tassin, L. Zhang, T. Koschny, A. P. Zhuravel, A. V. Ustinov,
S. M. Anlage, and C. M. Soukoulis, \textit{Classical Analogue of Electromagnetically Induced Transparency with a Metal-Superconductor Hybrid Metamaterial}, Phys. Rev. Lett. \textbf{107}, 043901 (2011).

\bibitem{Smith}
D. D. Smith, H. Chang, K. A. Fuller, A. T. Rosenberger, and R. W. Boyd, \textit{Coupled-resonator-induced transparency}, Phys. Rev. A \textbf{69}, 063804 (2004).

\bibitem{Tamayama}
Y. Tamayama, K. Yasui, T. Nakanishi, and M. Kitano, \textit{Electromagnetically induced transparency like transmission in a metamaterial composed of cut-wire pairs with indirect coupling}, Phys. Rev. B \textbf{89}, 075120 (2014).

\bibitem{Tassin2}
P. Tassin, L. Zhang, Th. Koschny, E. N. Economou, and C. M. Soukoulis, \textit{Low-Loss Metamaterials Based on Classical Electromagnetically Induced Transparency}, Phys. Rev. Lett. \textbf{102}, 053901 (2009).

\bibitem{Griffiths}
J. D. Griffiths, \textit{Introduction to Electrodynamics 4th ed.}, (Addison-Wesley 2012).

\bibitem{Gu}
J. Gu, R. Singh, X. Liu, X. Zhang, Y. Ma, S. Zhang, S. A. Maier, Z. Tian, A. K. Azad, H.-T. Chen, A. J. Taylor, J. Han, and W. Zhang, \textit{Active control of electromagnetically induced transparency analogue in terahertz metamaterials}, Nat. Commun. \textbf{3}, 1151 (2012).

\bibitem{Feng}
T. Feng, L. Wang, Y. Li, Y. Sun, and H. Lu, \textit{Voltage Control of Electromagnetically-Induced-Transparency-Like Effect in Metamaterials Based on Microstrip System}, Prog. Electromag. Res. Lett. \textbf{44}, 113 (2014).

\bibitem{Taflove}
A. Taflove, S. Hagnes, Computational Electrodynamics: The Finite-Difference Time-Domain Method 2nd ed, (Artech House, Inc., Norwood, MA, 2000).

\bibitem{Jahromi}
A. S. Jahromi, \textit{An extremely large group index via electromagnetically
induced transparency in metamaterials}, J. Europ. Opt. Soc. Rap. Public. \textbf{9}, 14048 (2014).

\end{thebibliography}
\end{document}